\documentclass[aps,floatfix,twocolumn,a4paper,noshowpacs, nofootinbib,
superscriptaddress,10pt]{revtex4}

\usepackage{graphicx,float}\usepackage{graphicx,float}
\usepackage[all]{xy}
\usepackage{amsmath,upgreek}
\usepackage{amssymb}
\usepackage{color}
\usepackage{epsfig}		
\usepackage{graphicx,epstopdf}
\usepackage{subfigure}
\usepackage{pdfpages}
\usepackage[colorlinks,hyperindex]{hyperref}

\definecolor{green1}{RGB}{0,128,0} 
\hypersetup{hidelinks,backref=true,pagebackref=true,hyperindex=true,colorlinks=true,breaklinks=true,urlcolor= blue}
\hypersetup{%
  colorlinks = true,
  linkcolor  = blue,
  citecolor = green1,
}
\usepackage{bookmark,textgreek}
\usepackage{hyperref,color,xcolor}
\hypersetup{hidelinks,hyperindex=true,colorlinks=true,breaklinks=true,urlcolor= blue}
\hypersetup{%
  colorlinks = true,
  linkcolor  = blue
}

\newcommand{\bes}{\begin{subequations}}
\newcommand{\ees}{\end{subequations}}
\def\beq{\begin{eqnarray}}
 
\def\eeq{\end{eqnarray}}
\def\be{\begin{equation}}
\def\ee{\end{equation}}

\usepackage{slashed}

\begin{document}

\title{Nucleons and higher spin baryon resonances: an AdS/QCD configurational entropic incursion}
  
\author{Luiz F.  Ferreira  }\email{luiz.faulhaber@ufabc.edu.br}
\affiliation{Federal University of ABC, Center of Mathematics,  Santo Andr\'e, Brazil}
\affiliation{Federal University of ABC, Center of Physics,  Santo Andr\'e, Brazil.}
\author{Roldao da Rocha}
\email{roldao.rocha@ufabc.edu.br}
\affiliation{Federal University of ABC, Center of Mathematics, Santo Andr\'e, Brazil}


\begin{abstract} 
 Families of $J^P=\frac12^+$ nucleons ($N(939)$, $N(1440)$, $N(1710)$, $N(1880)$, $N(2100)$, $N(2300)$) and $J^P=\frac32^-$ nucleons ($N(1520)$, $N(1700)$, $N(1875)$ and $N(2120)$) are scrutinized from the point of view of the configurational entropy (CE). 
The mass spectra of higher $J^P$ resonances in each one of these families are then obtained when configurational-entropic Regge trajectories, that relate the CE of nucleon families to both their $J^P$ spin and also to their experimental mass spectra, are interpolated. The   mass spectra of the next generation of nucleon resonances are then compared to already established baryonic states in PDG.  Besides the zero temperature case, the finite temperature analysis is also implemented.

\end{abstract}

\keywords{Configurational entropy; gauge/gravity correspondence;  AdS/QCD; baryons}

\maketitle

\section{Introduction}

The configurational entropy (CE) represents a measure of spatial correlations, in the very same sense of the pioneering  Shannon's information entropy. The fundamental interpretation of the CE consists of the limit to a lossless compression rate of information that is  
inherent to any physical system. The CE regards the minimum amount of bits necessary to encode a message \cite{Shannon:1948zz}. There is no code that takes less number of bits per symbol, on average, than the information entropy source. Codes closer to the information entropy are called optimal codes (or compression algorithms, in the case of noiseless sources), whose limit is the information entropy. 
This is the Shannon's coding-source theorem.  A physical system has maximal entropy when the distribution of probability  is uniform, since this is the highest unpredictable case. 
The CE is a measure of information concerning the spatial complexity of a system \cite{Gleiser:2011di,Gleiser:2012tu}, meaning the possible correlations among its parts along the spatial domain. 
The CE has been applied to non-linear scalar field models with spatially-localized energy  density \cite{Gleiser:2014ipa,Sowinski:2015cfa}. 
 Modes in a system corresponding to physical states with lower CE 
 have been shown to be more dominant and configurationally stable, being more detectable and observable, from the experimental point of view   \cite{Bernardini:2016hvx}.  

In what concerns quantum chromodynamics (QCD) as a reliable description of gluons and quarks (and their interactions), and other elementary particles as well, their intrinsic dynamical features make a difficult task of great complexity to investigate hadronic nuclear states and resonances. A particular approach to study non-perturbative aspects of QCD is the AdS/QCD model. It has been very successful to provide general properties of hadronic states,  hadronic mass spectra  and chiral symmetry breaking. These bottom-up models are implemented by considering deformations in the AdS space. For instance, the hard wall model consists in introducing an infra-red (IR) cut-off corresponding to the confinement scale \cite{Karch:2006pv}. On the other hand, in the case of the soft wall model is included a dilaton field acting as a smooth IR cut-off. 

In both these regimes, the CE provides new procedures  to figure out and unravel distinct physical phenomena.
The CE also consists of a powerful tool to probe, predict and corroborate 
to physical observables, including experimental and theoretical aspects. 
In fact, the CE has been playing important roles in scrutinizing  AdS/QCD (both hard wall and soft wall) models, studying relevant  properties in QCD and its phenomenology. The CE and the newly introduced concept of entropic Regge trajectories were used to investigate and predict the mass spectra of the next generation of higher spin particles of four light-flavour meson families  \cite{Bernardini:2018uuy,Ferreira:2019inu}, whereas an analog method was implemented in Ref. \cite{Ferreira:2019nkz}, to predict the mass spectra of higher spin tensor meson resonances, matching possible candidates  in PDG \cite{pdg1}. Previously, other aspects of the CE and mesonic states have been already studied in Ref. \cite{Bernardini:2016hvx}. In addition, the interplay between CE and the stability of scalar glueballs in  AdS/QCD was implemented in Ref. \cite{Bernardini:2016qit}. Ref. \cite{Braga:2017fsb} paved the scrutiny of the bottomonium and charmonium  production in the context of the CE in AdS/QCD. Phenomenological data show relative abundance and dominance of certain quarkonium states, whose intrinsic information is more compressed in Shannon's sense. Also, in the finite temperature case, it  identifies higher  phenomenological prevalence of lower $S$-wave resonances and lower masses quarkonia in Nature \cite{Braga:2018fyc}. Refs. \cite{Karapetyan:2016fai,Karapetyan:2017edu} utilized the CE to probe the dependence on the impact parameter of resonances production in $q\bar{q}$ scattering, inside the color-glass condensate (CGC) model. Besides, the production of diffractive mesonic resonances in AdS/QCD was proposed in Ref. \cite{Karapetyan:2018oye}, also in the CGC context. The CE was also employed to compute the pion cross section, in the holographic light-front wave function setup  \cite{Karapetyan:2018yhm}, whereas energy correlations in electron-positron annihilation were studied in \cite{Karapetyan:2019fst,Karapetyan:2019ran}. The anomalous dimension of the target gluon distribution and the nuclear CE were also shown to drive points of stability in particle collisions in LHC \cite{Karapetyan:2020yhs}. 
The CE setup to heavy-ion collisions was deployed in Ref. \cite{Ma:2018wtw}. Baryons were investigated under the CE apparatus, where pentaquarks were shown to have higher CE when compared to standard baryons \cite{Colangelo:2018mrt}. 
Ref. \cite{Braga:2020myi} studied quarkonium in finite density plasmas via the CE.

Not only the AdS/QCD regime of AdS/CFT has been explored from the point of view of the CE. 
 Stellar configurations and black holes were also explored via the CE  \cite{Gleiser:2015rwa}. Refs.  \cite{Braga:2016wzx,Braga:2019jqg}, besides showing that Hawking--Page phase transitions can be driven by the CE, demonstrated that the bigger the black holes, the more stable they are \cite{Lee:2018zmp,Lee:2017ero}.   The CE plays an important role on studying graviton Bose--Einstein condensates, as quantum portrait models of black holes. In fact, the Chandrasekhar collapse critical  density does correspond to a CE critical point    \cite{Casadio:2016aum,Fernandes-Silva:2019fez}. 
 Besides, the CE was shown to be an appropriate paradigm to study phase transitions as CE critical points \cite{Gleiser:2014ipa,Gleiser:2018kbq,Sowinski:2017hdw,Sowinski:2015cfa,Gleiser:2015rwa}. Topological  defects were also studied in Refs. \cite{Correa:2016pgr,Cruz:2019kwh,Bazeia:2018uyg}. The CE was also employed to derive the Higgs boson and the axion masses,  in effective theories  \cite{Alves:2014ksa,Alves:2017ljt}.

This paper is organized as follows: Sect. \ref{sec2} briefly introduces the soft wall AdS/QCD model,  presenting the nucleon resonances and the obtention of their mass spectra. In Sect. \ref{ce1} the CE is then computed for the $J^P=\frac12^+$ and $J^P=\frac32^-$ nucleon families. Hence, two types of configurational-entropic Regge trajectories are interpolated, relating the CE of the nucleon resonances to both their radial excitation and their  mass spectra. Therefore, the mass spectra of the next generation of higher spin nucleon resonances, for both $J^P=\frac12^+$ and $J^P=\frac32^-$ nucleon families are derived, from interpolation methods based on the already detected nucleon resonances in LHC.  The results are still compared to potential candidates in PDG \cite{pdg1}. The finite temperature case is also investigated. Sect. \ref{sec4} is devoted to a further analysis, discussion and compilation of the main results and their physical consequences.

\section{Holographic AdS/QCD model and baryons}
\label{sec2}
The description of $N$ baryons with spin $J^P=\frac12^+$ in the AdS/QCD soft wall model can be accomplished by using different approaches \cite{deTeramond:2005su,Colangelo:2008us,Hong:2006ta,Forkel:2007cm,Gutsche:2011vb,Li:2013lfa,Fang:2016uer,FolcoCapossoli:2019imm,Gutsche:2019pls}. We choose the method employed in Refs. \cite{Gutsche:2019blp,Gutsche:2019pls}  to describe fermions, both at zero and finite temperature,  in the AdS/QCD soft wall model. Due the fact that the solutions in Refs. \cite{Gutsche:2019blp,Gutsche:2019pls} are degenerate, for any value of half-integer $J^P$,  and hold for  hadrons composed of angular orbital  momentum $\ell$ and radial quantum number $n$, these solutions will be also applied for the  $N$ baryon family with  $J^P=\frac32-$. 

To start, one considers the AdS--Schwarzschild metric in conformal coordinates, 
\begin{equation}\label{metricSCh}
ds^2=\frac{R^2}{z^2}\left(f_{T}(z)dt^2-d\vec{x}^2-\frac{dz^2}{f_{T}(z)} \right),
\end{equation}
with $f_{T}(z)=1-z^4/z_{h}^{4}$,  where $z_h$ is proportional to the inverse of the  AdS--Schwarzschild black brane temperature as $T=1/(\pi z_h)$. Besides the quadratic  dilaton field $\phi(z)=k^2z^2$  usually employed in the AdS-QCD soft wall model, a thermal pre-factor, $e^{-\lambda_T(z)}$, is also introduced, where 
\begin{equation}\label{thermalprefactor}
\lambda_T(z)=\upalpha\frac{z^2}{z_{h}^{2}}+\gamma \frac{z^4}{z_{h}^{4}}+\upxi \frac{z^6}{z_{h}^{6}}\,,
\end{equation}
where the dimensionless parameters $\gamma$ and $\upxi$  have been fixed in Ref. \cite{Gutsche:2019blp} to guarantee  gauge invariance and massless ground-state pseudoscalar mesons ($\pi$, $K$, $\eta$) in the chiral limit, as well as to suppress the sixth power of the radial coordinate in the holographic potential. Precisely, $\upgamma=\frac{J(J-3)+3}{5}$ and $\upxi=\frac{2}{5}$. The term $\upalpha z^2/z^{2}_{h}$ is a small perturbative correction to the dilaton $\phi$, where the parameter $\upalpha$ is related with the restoration of chiral symmetry at critical temperature. It is useful to relate the Regge--Wheeler (RW) tortoise coordinate, $r$, to the holographic coordinate, $z$, via the substitution: 
\begin{equation}\label{RW}
r\!=\!\int\! \frac{dz}{f(z)}\!=\!\frac{z_h}{2}\left[\frac{1}{2}\log \left(\frac{1-z/z_h}{1+z/z_h}\right)-\arctan \left( \frac{z}{z_h} \right)\right]\,.
\end{equation}
One expands the $r$ coordinate at low temperature ($z_h \to \infty$) and restrict, to the leading order (LO) and to the next  leading (NLO), terms in the expansion of $z$ in powers of $r$  \cite{Gutsche:2019blp,Gutsche:2019pls},
\begin{equation}\label{expz}
z=r\left[1-\frac{t^{4}_{r}}{5}+\mathcal{O}(t^{8}_{r})\right], \,\,\,\,\,\, t_{r}=\frac{r}{z_h}\,.
\end{equation}
Hence, using the expansion in Eq. (\ref{expz}), the metric \eqref{metricSCh} can be  replaced by:
\begin{equation}\label{metricr}
ds^2=e^{2A(r)} f_{T}^{3/5}(r)\left(dt^2-dr^2-\frac{d\vec{x}^2}{f_{T}(r)} \right).
\end{equation}
where the warp factor reads $A(z)=\log{\left(R/r\right)}$  and $f_{T}(r)=1-r^4/z_{h}^{4}$. Considering the small correction to the dilaton, one defines a thermal dilaton as 
\begin{eqnarray}\label{thermaldilaton}
\phi_{T}(r)&=&k^2_{T}r^2=k^2r^2(1+\rho_{T}),
\end{eqnarray}
where 
\begin{eqnarray}\label{thermaldilaton1}
 \rho_{T}\!=\!\left( \frac{9\upalpha \pi^2}{16}\!+\!\delta_{T_1}\right)\frac{T^2}{12F^2}\!+\!\delta_{T_{2}}\left(\frac{T^2}{12F^2}\right)\!+\!\mathcal{O}(T^6),
\end{eqnarray}
with parameters, matching phenomenological data,  fixed as in  $\upalpha=0$,  $\delta_{T_1}=-\frac83$, $\delta_{T_2}=-\frac49$  and $F=87 $ MeV \cite{Gutsche:2019blp}. 

The action for  fermionic field $\Psi$ takes  the following form  at finite temperature:
\begin{eqnarray}\label{actionBaryon}
&&S_{B}=\int d^4x dr \sqrt{g}\bar{\Psi}(x,r,T)\hat{\mathcal{D}}_{\pm}(r)\Psi(x,r,T), \end{eqnarray}
where the Dirac operator reads
\begin{eqnarray}\label{actionBaryon1}
\!\hat{\mathcal{D}}_{\pm}(r)\!=\!\frac{i}{2}\Upgamma^{M}\!\left[ \partial_{M}\!-\!\omega_{M}^{ab}[\Upgamma_{a},\Upgamma_{b}]\right] \!-\!\left[ m_5(r,T)\!+\!V_{\Psi}(r,T)\right],
\end{eqnarray}
for $\Upgamma^{a}=(\upgamma^{\mu},-i\upgamma^{5})$ denoting the Dirac matrices, $\upgamma^{5}=i\upgamma^{0}\upgamma^{1}\upgamma^{2}\upgamma^{3}$ is the volume element, $\epsilon^{a}_{M}=R \delta^{a}_{M}/z$ denotes the  tetrad and  $\omega^{ab}_{M}=\frac{1}{4rf^{1/5}(r)}(\delta^{[a}_{r}\delta^{b]}_{M})$ denotes the   spin connection. The quantity $ m_5(r,T)=m_5/f^{3/10}(r)$ is the temperature associated with the bulk  fermion mass in AdS space. Here one considers  $m_{5}=\bar{\tau}-2$ \cite{Forkel:2007cm}, with  $\bar{\tau}$ corresponding to the twist dimension of the baryon operator at the boundary, which is related with the orbital angular momentum \cite{deTeramond:2005su,Brodsky:2007hb,Brodsky:2010ur} as $\bar{\tau}=\ell+3$. For instance, the $J^{P} =\frac12^{+}$ states are related with  $\ell = 0$ and   $J^{P} =\frac12^{-}$ states with $\ell = 1$. The dilaton temperature potential reads \cite{Gutsche:2019pls}
\begin{eqnarray}
 V_{\Psi}(r)=\frac{\phi_{T}(r,T)}{f^{3/10}(r)}\,.
\end{eqnarray}
The baryon field can be split into left-handed, $L$, and right-handed, $R$, components:
\begin{eqnarray}\label{LR}
\Psi(x,z)=\Psi_{L}(x,z)+\Psi_R(x,z), 
\end{eqnarray}
where the chiral spinor components are given by 
\begin{eqnarray}\label{LR}
\Psi_{L/R}(x,z)=\frac{1\mp \upgamma^{5}}{2}\Psi(x,z).   
\end{eqnarray}

One can perform a Kaluza--Klein expansion of the 4D transverse components of the  field, 
\begin{equation}\label{Kaluza}
\Psi^{L/R}(x,r,T)= \sum_{n} \Psi^{L/R}(x)\tau^{L/R}_{n/2}(r,T),
\end{equation}
with  $\Psi^{L/R}(x)$  denoting  the tower of Kaluza-Klein (KK) modes and $n$ is the radial quantum number. Therefore, replacing 
\beq\label{taul}
\tau^{L/R}(r,T)=e^{-\frac32 A(r)}\upchi^{L/R} (r,T),
\eeq in the equation of motion obtained from the action (\ref{actionBaryon}), yields a Schr\"odinger-type equation of motion for $\upchi (r,T)$, given by 
\begin{equation}\label{SchrodingerEOMS}
\left[-\partial^{2}_{r}+U_{L/R}(r,T)  \right] \upchi^{L/R} (r,T)=M^2_{n}\upchi^{L/R} (r,T).
\end{equation}
The effective potential $U_{L/R}(r,T)$ at finite temperature can be split into two parts: 
\begin{eqnarray}\label{potential}
U_{L/R}(r,T)&=&\Lambda_{L/R}(r)+\Omega_{L/R}(r,T),
\end{eqnarray}
where 
\begin{eqnarray}\label{potential1}
 \!\!\Lambda_{L/R}(r)\!\!\!&\!\!\!=\!\!\!&\!\!k^4r^2\!+\!2k^2\left(\!\ell\!+\!1\!\mp\! \frac{1}{2}\right)\!+\!\frac{(\ell\!+\!1)(\ell\!+\!1\!\pm\! 1)}{r^2}, \nonumber\\   \Omega_{L/R}(r,T)&=&2 \rho_{T} k^2\left(k^2r^2 \right)\,.
\end{eqnarray}
The normalizable solutions of Eq. (\ref{SchrodingerEOMS}), for the baryons at finite temperature, read 
\begin{eqnarray}\label{solL}
\!\!\!\!\!\upchi_{n}^L(r)\!=\!\sqrt{\frac{2 n!}{\Gamma(n\!+\!\ell\!+\!\frac52)}}e^{-k_{T}^2r^2/2}k_{T}^{\ell\!+\!\frac52}r^{\ell\!+\!2}L^{\ell\!+\!\frac32}_{n}(k_{T}^2r^2),\\
\label{solR}
\!\!\!\!\!\upchi_{n}^R(r)\!=\!\sqrt{\frac{2n!}{\Gamma(n\!+\!\ell\!+\!\frac32)}}e^{-k_{T}^2r^2/2}k_{T}^{\ell\!+\!\frac32}r^{\ell\!+\!1}L^{\ell\!+\!\frac12}_{n}(k_{T}^2r^2),
\end{eqnarray}
where $\Gamma(p)$ is the gamma function, $L_n$ is the Laguerre function and the mass spectrum is specified by 
\begin{equation}\label{Spectrum}
M^2_{n}=4k_{T}^2\left(n+\ell+\frac32\right)\,.
\end{equation}

The eigenfunctions (\ref{solL}, \ref{solR}) are normalized according to 
\begin{equation}\label{normal}
\int^{\infty}_{0}dz\upchi_{m}^{L/R}(r,T)\upchi_{n}^{L/R}(r,T)=\delta_{mn}\,.
\end{equation}
In the zero temperature limit,  $T \rightarrow 0$, the solution of Eq. (\ref{SchrodingerEOMS})  reads
\begin{eqnarray}\label{solLzero}
\!\!\!\!\!\upchi_{n}^L(r)\!=\!\sqrt{\frac{2n!}{\Gamma(n\!+\!\ell\!+\!\frac52)}}e^{-k^2r^2/2}k^{\ell\!+\!\frac52}r^{\ell\!+\!2}L^{\ell\!+\!\frac32}_{n}(k^2r^2),\\
\label{solRzero}
\!\!\!\!\!\upchi_{n}^R(r)\!=\!\sqrt{\frac{2n!}{\Gamma(n\!+\!\ell\!+\!\frac32)}}e^{-k^2r^2/2}k^{\ell\!+\!\frac32}r^{\ell\!+\!1}L^{\ell\!+\!\frac12}_{n}(k^2r^2),
\end{eqnarray}
which corresponds to the mass spectrum
\begin{equation}\label{Spectrum2}
M^2_{n}=4k^2\left(n+\ell+\frac32\right)\,.
\end{equation}
It is worth to emphasize that in the infrared regime of the $r$ coordinate, the tau functions in Eq. (\ref{taul}) scale as $\tau^{L}(r,T)\sim r^{\frac92+\ell}$ and $\tau^{R}(r,T)\sim r^{\frac72+\ell}$, whereas at the ultraviolet regime these functions vanish, exhibiting confinement \cite{Gutsche:2011vb}.

\begin{figure}[h!]
\begin{center}
\includegraphics[scale=0.45]{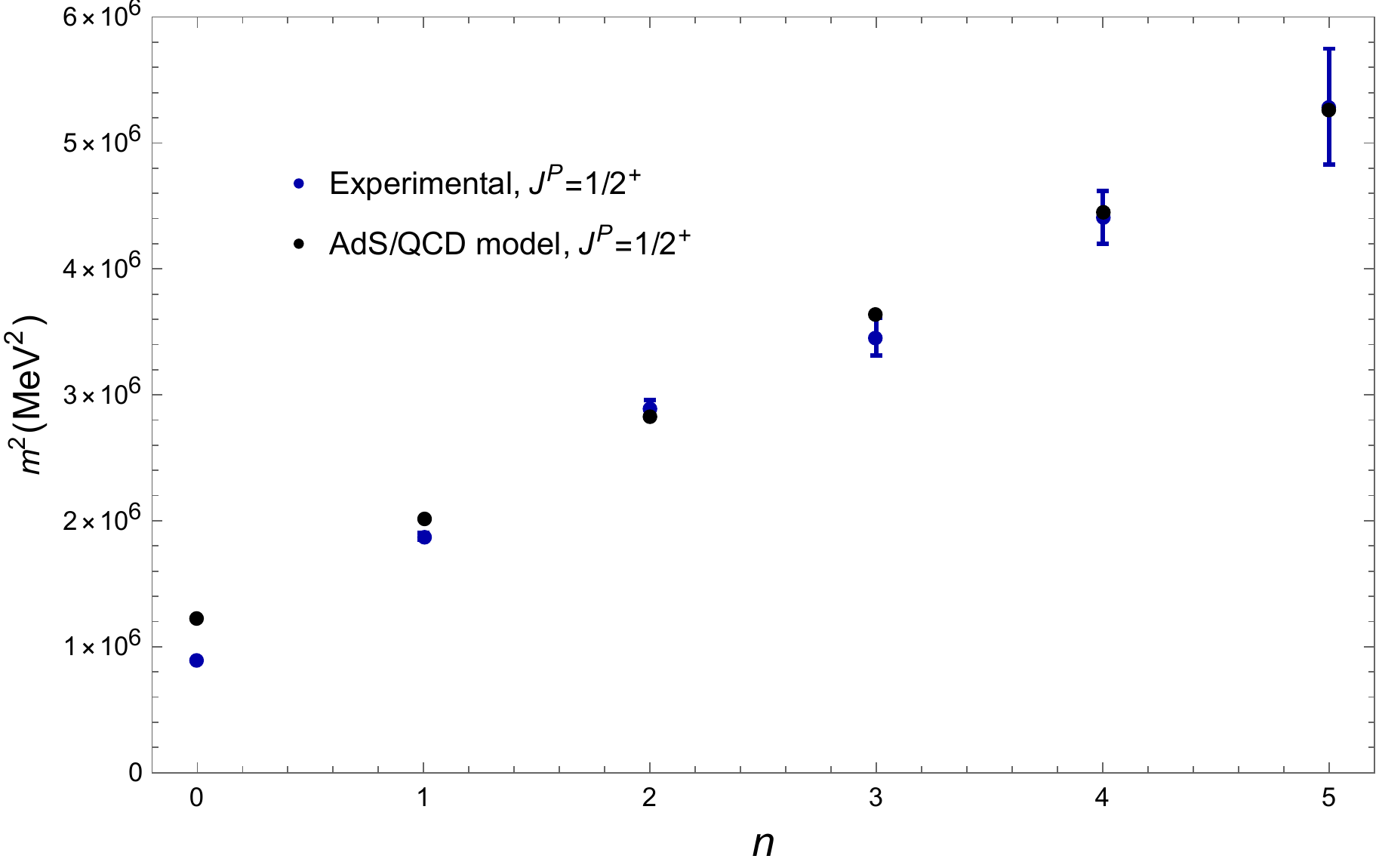}
\end{center}
\caption{Mass spectra of nucleon resonances with $J^P=\frac12^+$ obtained from the AdS/QCD model and  experimental values \cite{pdg1}.}
\label{nuc2}
\end{figure}

First, the nucleon family with $J^P=\frac12^+$ will be approached.
 To obtain the best fit with the experimental data at zero temperature, one fixes the parameter  $k=0.45$ GeV for the $J^P = \frac12^+$ nucleon resonances. For comparison,  the experimental data from \cite{pdg1} are presented in the third column and the soft wall result (\ref{Spectrum}) in the fourth column of Table \ref{Nucleonmasses}. 
 \begin{table}[h]
\begin{center}--------- $J^P=\frac12^+$ nucleon mass spectra (MeV) ---------\medbreak
\begin{tabular}{||c|c||c|c||}
\hline\hline
        $n$ & Nucleon ($J^P=\frac12^+$)       & Experimental &  AdS/QCD (Soft wall)  \\\hline\hline
\hline
         \textcolor{black}{0} & $N(939)$    & $939.49 \pm 0.05$     & 1102.27             \\\hline
       \textcolor{black}{1} &   $N(1440)$ & $1370\pm10$    & 1423.02 \\\hline
         \textcolor{black}{2}& $N(1710)$           & $1700\pm20$     & 1683.75            \\\hline
         \textcolor{black}{3} & $N(1880)$         & $1860\pm40$       & 1909.19 \\\hline
         \textcolor{black}{4} & $N(2100)$           & $2100\pm50$              & 2110.69   \\\hline
         \textcolor{black}{5} & $N(2300)^*$          & $2300^{+40  +109}_{-30-0}$     & 2294.56               \\
\hline\hline
\end{tabular}
\caption{Mass spectra of the $J^P=\frac12^+$ nucleon family resonances. The particle with an asterisk is left out the summary table in PDG \cite{pdg1}. }
 \label{Nucleonmasses}
\end{center}
\end{table}
This makes possible to illustrate the mass spectra (\ref{Spectrum}) of  nucleon resonances with $J^P=\frac12^+$  and realize that they match experimental values  \cite{pdg1}. This is accomplished in Fig.  \ref{nuc2}. 
 
\begin{figure}[h!]
\begin{center}
\includegraphics[scale=0.5]{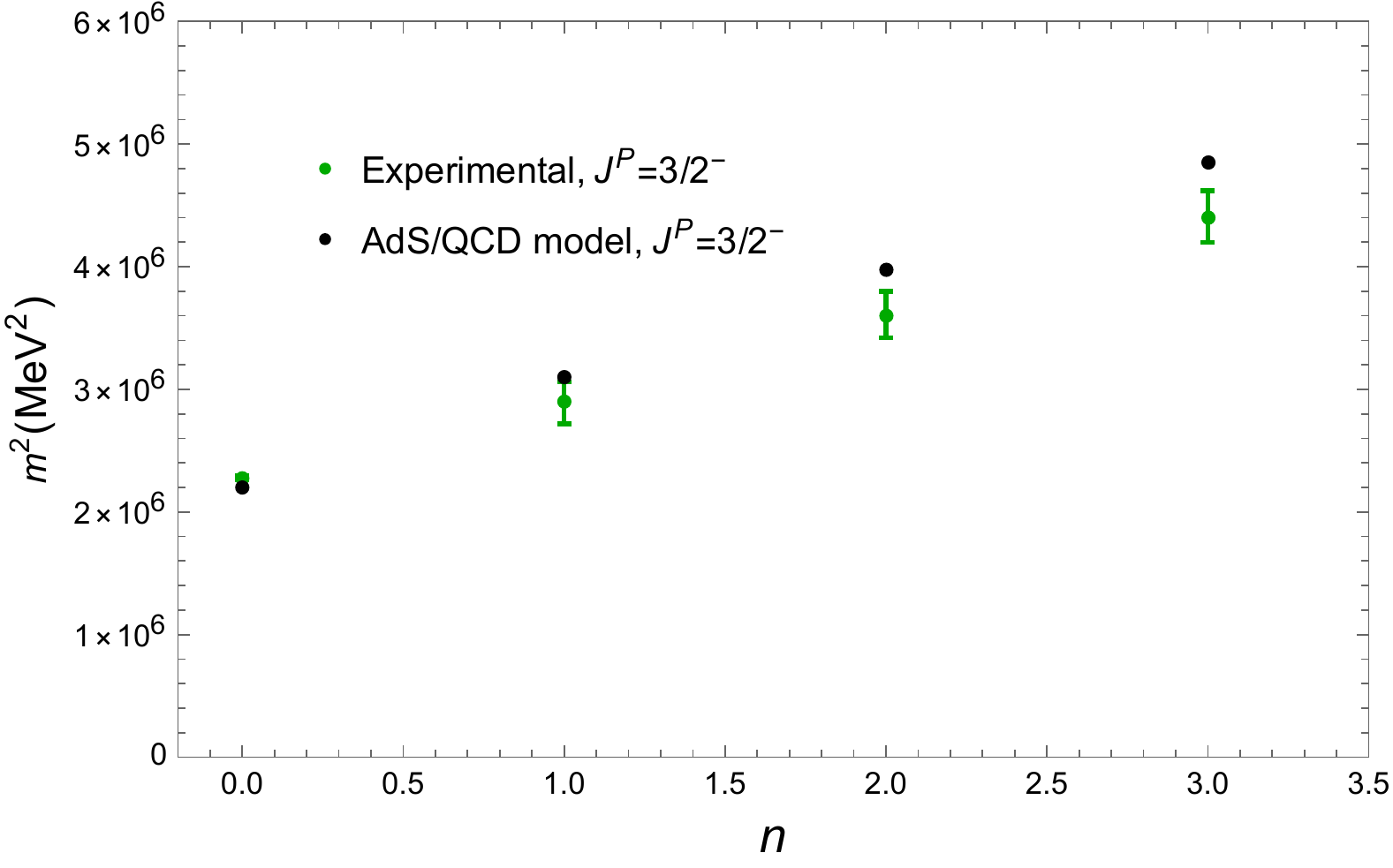}
\end{center}
\caption{ Mass spectra for  nucleon resonances  with $J^P=\frac32^-$ obtained from the AdS/QCD model and the experimental values \cite{pdg1}. }
\label{massDelta1}
\end{figure}

Now, nucleon resonances with  $J^P = \frac32^-$ will be investigated. They consist of excited states of nucleon particles, often corresponding to one of the quarks having a flipped spin state, or with different orbital angular momentum when the particle decays.   The parameter of the dilaton is fixed as $k=0.47$ GeV, having the best fit with the experimental results as one can see in Fig. \ref{massDelta1}.  Table \ref{Deltamasses} shows the  mass spectra in the soft wall model for baryons with $J^P = \frac32^-$ and their  experimental results \cite{pdg1}. 
\begin{table}[h!]
\begin{center}--------- $J^P=\frac32^-$ Nucleon mass spectra (MeV) ---------\medbreak
\begin{tabular}{||c|c||c|c||}
\hline\hline
        $n$ & Nucleon  ($J^P=\frac32^{-})$       & Experimental &  AdS/QCD (Soft wall)  \\\hline\hline
\hline
         \textcolor{black}{0} &  $N(1520)$    & $1510\pm5$   & 1486.27            \\\hline
       \textcolor{black}{1} &   $N(1700) $ & $1700\pm50$    &  1758.57 \\\hline
         \textcolor{black}{2}& $N(1875)$           & $1900\pm50$      & 1994.04           \\\hline
         \textcolor{black}{3}& $N(2120)$           &    $2100\pm50$  & 2204.49           \\\hline
         \hline\hline
\end{tabular}
\caption{Mass spectra of $J^P=\frac32^{-}$ nucleon family resonances. }
 \label{Deltamasses}
\end{center}
\end{table}

\newpage
\section{Configurational entropy and Shannon information entropy}\label{ce1}
 When one takes into account some  distribution of probability $p_a$, the CE is defined, for discrete systems, by (minus) the sum $\sum_i p_i\,\log(p_i)$ \cite{Gleiser:2012tu,Sowinski:2015cfa}. When $N$ modes, constituting a discrete physical system,  are described by a uniform 
distribution of probability, namely, $p_i = 1/N$, the CE has a maximal value equal to $\log N$. 
The differential CE regards the continuous limit of the CE \cite{Gleiser:2018kbq,Bernardini:2016hvx}. 
The CE, underlying any physical system, has the main pillar on the spatial correlations among the parts of the system. More precisely, 
the central structure of the CE resides on the correlation of the fluctuations of some $\uprho(x)$  scalar field, which is a localized and Lebesgue integrable function describing the system. Given a vector  $\mathtt{r}\in\mathbb{R}^d$, the two-point correlation function is given by  $G(\mathtt{r}) = { \int_{\mathbb{R}^d}\uprho(\tilde{\mathtt{r}})\uprho(\mathtt{r}+\tilde{\mathtt{r}})d\tilde{\mathtt{r}}}$. The CE is nothing else than the Shannon's information theory involving correlations, if one establishes a distribution of probability using the two-point correlation function. 
To accomplish it, the so called power spectrum, $P(\mathtt{k})$, where $\mathtt{k}$ is the wave vector, is defined to be the  Fourier transform of the two-point correlation function \cite{Braga:2018fyc}.  The convolution and the Plancherel theorems yield 
$P(\mathtt{k}) \sim \| \int_{\mathbb{R}^d}\uprho({\mathtt{r}})e^{i\mathtt{k}\cdot \mathtt{r}}\,d{\mathtt{r}}\|^2$. When one takes the Fourier transform 
\beq\label{fou}
\uprho(\mathtt{k}) = \frac{1}{(2\pi)^{p/2}}\int_{\mathbb{R}^d}\uprho(\mathtt{r})e^{-i\mathtt{k}\cdot \mathtt{r}}\,d{\mathtt{r}},\eeq 
then the correlation distribution of probability, called as the modal fraction, reads   
\cite{Gleiser:2012tu}
\begin{eqnarray}
\varrho(\mathtt{k}) = \frac{|\uprho(\mathtt{k})|^{2}}{ \int_{\mathbb{R}^d}  |\uprho(\mathtt{k})|^{2}d^d\mathtt{k}}.\label{modalf}
\end{eqnarray} 
Therefore, the CE reads 
\begin{eqnarray}
{\rm CE}_\uprho= - \int_{\mathbb{R}^d}{\varrho_\star}(\mathtt{k})\log {\varrho_\star}(\mathtt{k})\, d^dk\,,
\label{confige}
\end{eqnarray}
where $\varrho_\star(\mathtt{k})=\varrho(\mathtt{k})/\varrho_{\rm max}(\mathtt{k})$
The expression of the CE is similar to the Gibbs entropy, which is defined for some statistical ensemble, of microstates having each one probability $p_i =
e^{-E_i/k_{\rm b}T}/\sum_i e^{-E_i/k_{\rm b}T}$, where $E_i$ stands for the energy of the $i^{\rm th}$ microstate and $k_{\rm b}$ represents the Boltzmann constant. The interplay between the CE and statistical mechanics was paved in \cite{Bernardini:2016hvx}.

Both for the  $J^P=\frac12^+$ and the $J^P=\frac32^-$ nucleon families, given the Lagrangian $L$ ruling the system in Eq. (\ref{actionBaryon}), the energy-momentum tensor is, in general, given by 
 \begin{equation}
 \!\!\!\!\!\!\!\!T^{\mu\nu}\!=\!  \frac{2}{\sqrt{ g }}\!\! \left[ \frac{\partial (\sqrt{g}{L})}{\partial g_{\mu\nu} }\!-\!\partial_{ x^\alpha }  \frac{\partial (\sqrt{g} {L})}{\partial \left(\!\frac{\partial g_{\mu\nu} }{\partial x^\alpha}\!\right) }
  \right].
  \label{em1}
 \end{equation} 
 \noindent  
 The $\uprho(z)$ energy density corresponds to the $T_{00}(z) $ component of (\ref{em1}), emulating Eq. (47) of Ref. \cite{Colangelo:2018mrt} for our specific case, 
 as
 \beq\label{t00}
\uprho(r)= T_{00}(r)=\frac{M_n^2}{r}\left[\left(\upchi_{n}^L(r)\right)^2+\left(\upchi_{n}^R(r)\right)^2\right]\,.
 \eeq
 Hence, the energy density (\ref{fou}) in momentum space the modal fraction (\ref{modalf}) and the CE (\ref{confige}) can be numerically computed. 

The CE     of the $J^P=\frac12^+$ nucleons $N(939)$, $N(1440)$, $N(1710)$, $N(1880)$, $N(2100)$, $N(2300)$, as a function of the $n$ quantum number, is displayed in the third column of Table \ref{CES}, for $n=0,1,\ldots,5$. For higher values of $n$, the CE 
will be computed through the interpolation method, to be described in what follows. 
  \begin{table}[h]
\begin{center}\medbreak
\begin{tabular}{||cc||c|c|||c||c||}
\hline\hline
  &   $n$ & $N$ ($J^P=\frac12^{+}$) &~CE & $N$  ($J^P=\frac32^{-}$) &~CE \\\hline\hline
     &  \, 0 \,&\, $N(939)$& $~ 0.94981~$ & $N(1520) $  &~$0.99604$~ \\\hline
     \,&\,   1 \,&\, $N(1440)$&$~1.06402~$ & $N(1700)$ &~$1.14581$~ \\\hline
     \,&\,   2 \,&\, $N(1710)$&$~1.15043~$ & $N(1875)$ &~$1.22903$~\\\hline
     \,&   3\, &\, $N(1880)$&$~1.21481~$  & $N(2120)$  &~$1.28816$~ \\\hline
     \,&\,   4 \,&\, $N(2100)$&$~1.26504~$ & $X_{5}$ &~$1.38194$~ \\\hline
     \,&\,   5 \,&\, $N(2300)$&$~1.30562~$ & $X_{6}$ &~$1.59038$~ \\\hline
     \,&\,  6 \,&\, $N_7 $\,&$ ~1.35258~$ & $X_{7}$  &~$\,2.05683$ ~ \\\hline
     \,&\,7\,&$\,N_8$ &~$1.41702$~  &  & \\\hline
     \,&\,8\,&$\,N_9$&~$1.51519$~ &  & \\\hline
\hline
\end{tabular}\caption{The CE for the nucleon family  as a function of their radial excitation $n$, in the soft wall model. The 2${}^{\rm nd}$ column displays the nucleon family with spin $J^P=\frac12^{+}$ (and the higher spin resonances $N_7, N_8, N_9$), whereas the 3${}^{\rm rd}$ column shows their respective CE; the 4${}^{\rm th}$ column shows the nucleon family with spin $J^P=\frac32^{-}$ (and the higher spin resonances $X_5, X_6, X_7$) and the 5${}^{\rm th}$ column illustrates their respective CE. }
\label{CES}
\end{center}
\end{table}  

     \begin{figure}[H]
\centering
\includegraphics[width=8.5cm]{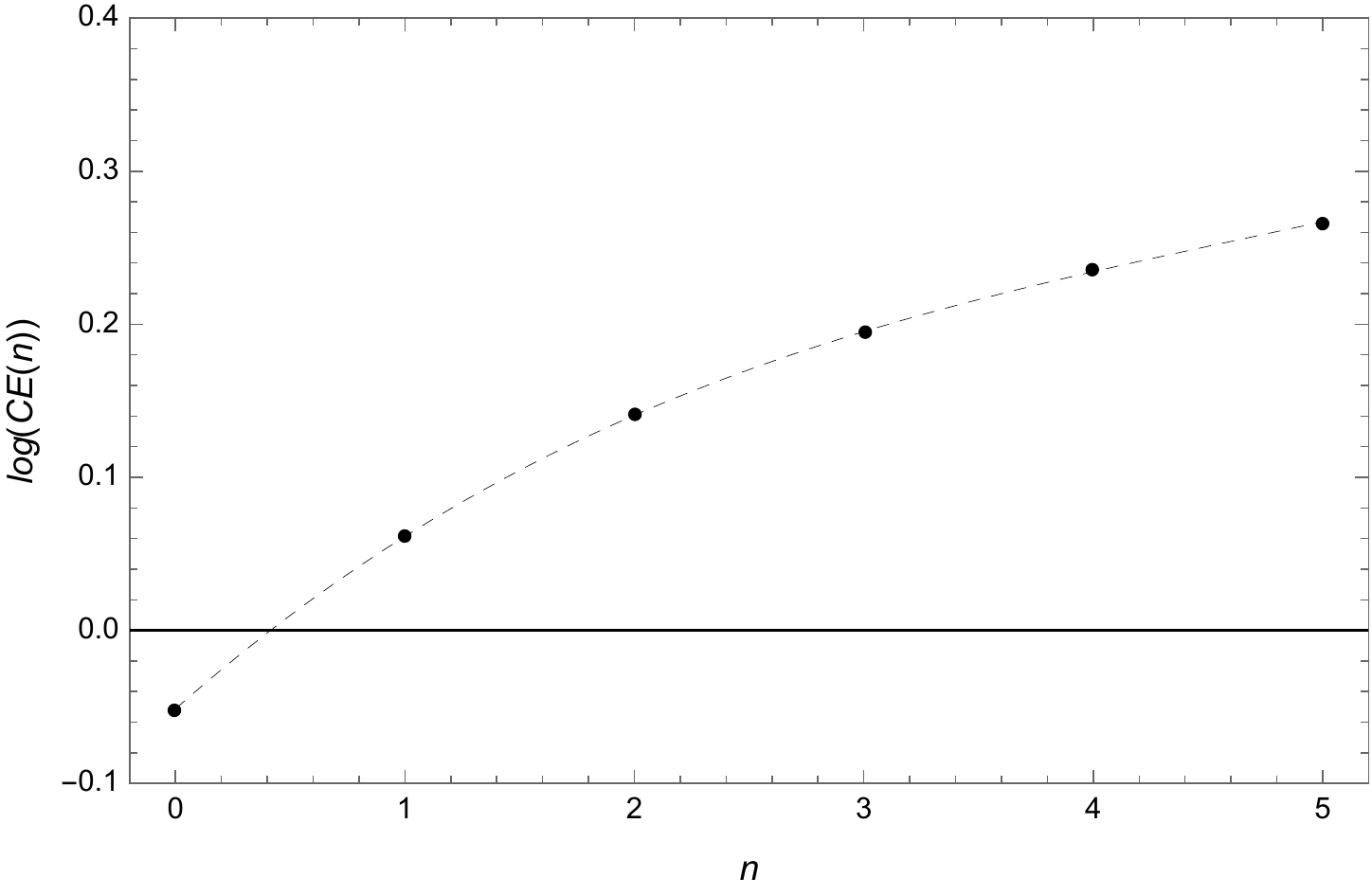}
\caption{$\log$(CE) of the $J^P=\frac12^+$ nucleons as a function of the $n$ quantum number.}
\label{kaon1}
\end{figure}
\noindent For the $J^P=\frac12^+$ nucleon family, the  configurational-entropic Regge trajectory, 
relating $\log$(CE) to the $n$ quantum number, is the interpolation dashed line in Fig. \ref{kaon1}. It has the 
explicit expression
\begin{eqnarray}
 {\rm CE}(n) &=&1.06283\times10^{-3} \; n^{3} -0.01712 \;n^2  \nonumber\\&&0.13025 n+0.94983.
 \label{itp0}
   \end{eqnarray} We choose to use the third power in the interpolation, as it is sufficient to delimit within $\sim0.17\%$ the standard deviation.

A second type of configurational-entropic Regge trajectory, interpolating and relating the CE to (squared) mass spectra of the nucleon family $J^P=\frac12^+$, is shown as the dashed curve in Fig. \ref{f62},
\begin{figure}[H]
\label{f62}
\begin{center}
\includegraphics[scale=0.55]{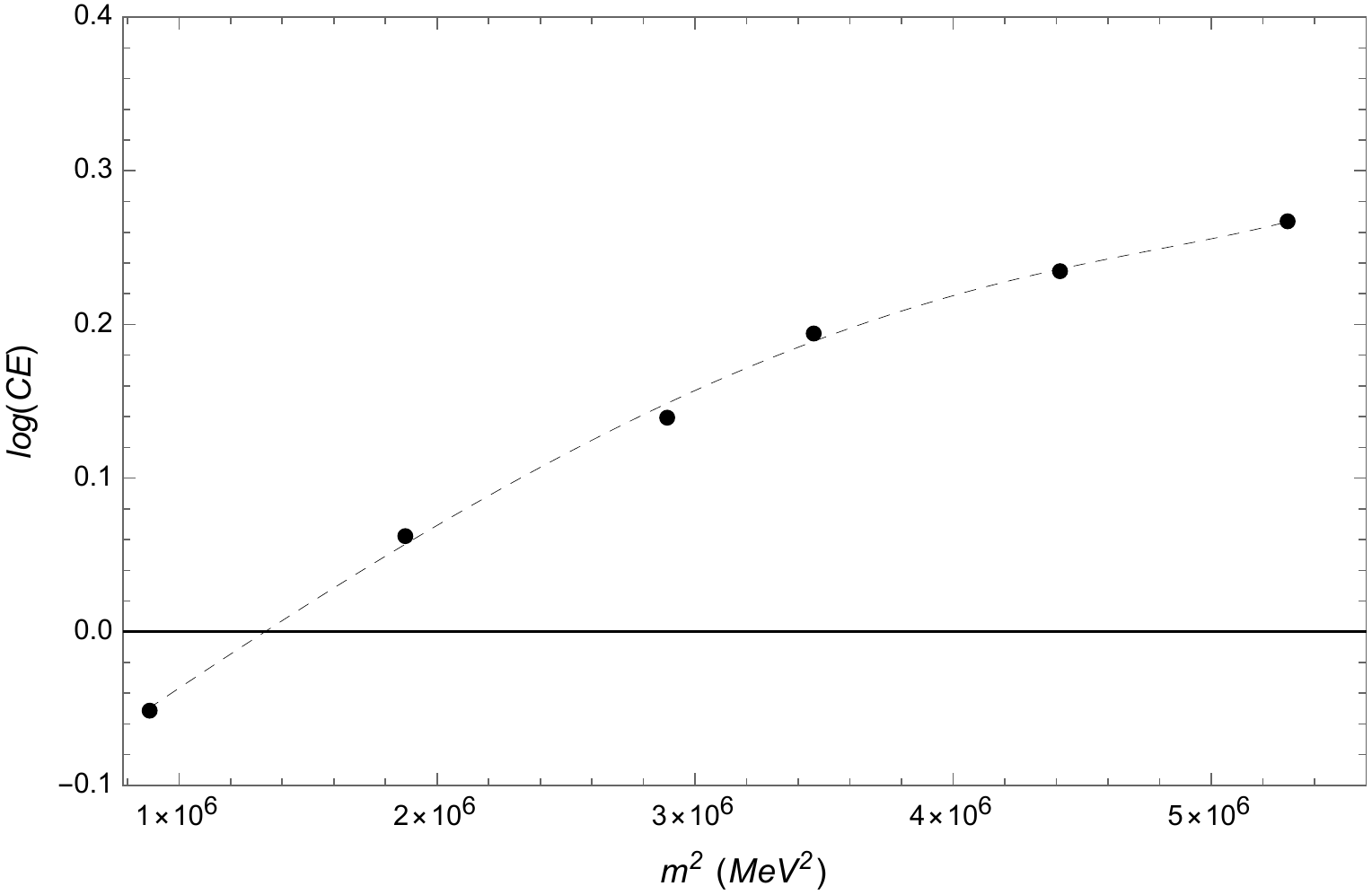}
\end{center}
\caption{ The CE for the nucleon family $J^P=\frac12^+$ information-entropic Regge trajectory, with respect to the squared mass spectra of the nucleon family $J^P=\frac12^+$.}\label{f62}
\end{figure} \noindent {\color{black}{The dashed curve in Figs. \ref{kaon1} and \ref{f62}, respectively correspond to 
Eqs. (\ref{itp0}) and (\ref{itq0}).}}
 For the plot in Fig.  \ref{f62}, the configurational-entropic Regge trajectory has the following 
form:
\begin{eqnarray}\label{itq0}
\!\!\! \log{\rm CE}(m) \!&=\!&\! 3.09873\times 10^{-50} m^{14} -1.16011 \times 10^{-14} m^6  \nonumber\\&&1.42784 \times 10^{-7} m^2-0.167903,    \end{eqnarray} within $1.5\%$ standard deviation. {\color{black}{It is worth to emphasize that in Eq. (\ref{itq1}) $m$ refers to the nucleon mass.}}

 The mass spectra for the $N_{7}, N_{8}$ and $N_{9}$ elements can be easily inferred, by employing Eqs. (\ref{itp0}, \ref{itq0}). In fact, for $n=6$, Eq. (\ref{itp0}) yields 
${\rm CE} = 1.35258$. Then substituting this value in the CE Regge trajectory (\ref{itq0}), after solving the subsequent equation yields the mass 
\beq
m_{N_7}=2880\pm 43 \;{\rm MeV}.
\eeq  
It is worth to mention that this value of mass complies with the order of the one for the  
$\Lambda_c(2940)^+$ baryon, also with $J^P=\frac32^-$, with mass $2939.6^{+1.3}_{-1.5}$, detected in LHC \cite{pdg1}.

Similarly, for the next member $N_{8}$ of the $J^P=\frac12^+$ nucleon family, substituting $n=7$ into Eq. (\ref{itp0}) implies that the corresponding CE has value $1.41702$. Hence, replacing this value into 
the CE Regge trajectory (\ref{itq0}) yields \beq
m_{N_{8}}=3062\pm47\; {\rm MeV}.\label{mn8}
\eeq 
The 
$N_8$ nucleon resonance matches the mass $3055.9\pm0.4$ of the $\Xi_c(3055)$
 baryon \cite{pdg1}. As other properties, like $J^P$ and the isospin are still undetermined for $\Xi_c(3055)$, where only 894 events have been detected in LHC, 
the $\Xi_c(3055)$ baryon might emulate  a potential candidate to describe the $N_8$ nucleon resonance with mass (\ref{mn8}),  just in the case where $J^P=\frac32^-$ and the isospin $\frac12$  match the ones for $\Xi_c(3055)$.

Besides, the same can be accomplished for the $N_{9}$,   $J^P=\frac12^+$ nucleon, when Eq. (\ref{itp0}) implies that ${\rm CE}_{N_{9}} = 1.51519$. When replaced into Eq. (\ref{itp0}), it produces the mass 
\beq\label{mn9} 
m_{N_{9}}=3169\pm51\; {\rm MeV}. 
\eeq
Once the $N_9$ nucleon resonance mass \eqref{mn9} 
has agreement to the mass $3122.9\pm1.6$ of the $\Xi_c(3123)$
 baryon \cite{pdg1}, with $J^P$ and the isospin still undetermined, 
the $\Xi_c(3123)$ baryon might consist of a candidate to describe the $N_9$ nucleon resonance with mass (\ref{mn9}). In the same way, the $\Omega_c(3120)^0$, with mass $3119.1\pm1.5$ might also be described by the $N_9$ nucleon, if $J^P$ and the isospin will be shown to match, indeed.

\begin{figure}[H]
\centering
\includegraphics[width=8cm]{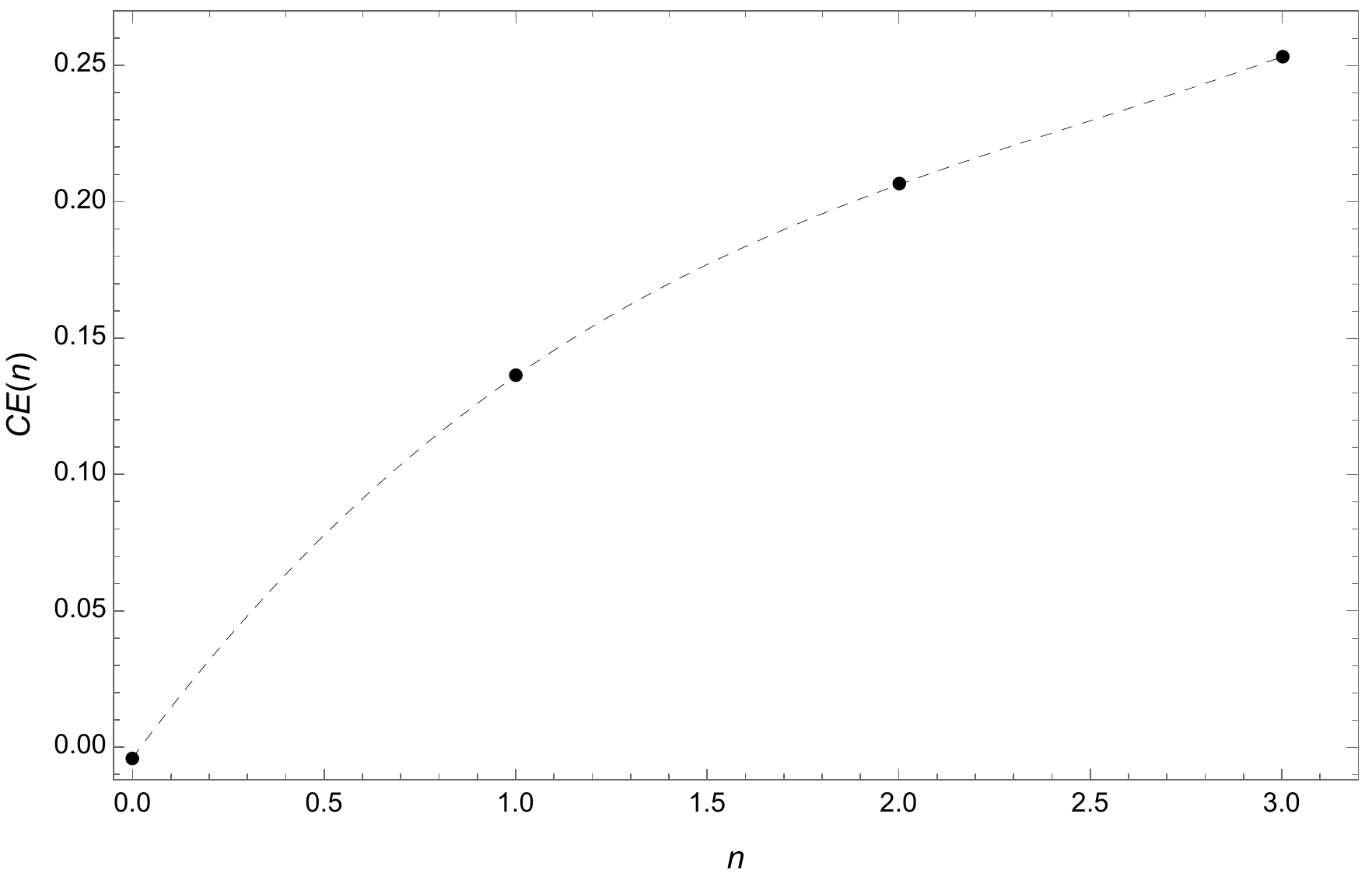}
\caption{CE of the $J^P=\frac32^-$ nucleons as a function of the $n$ quantum number.}
\label{kaon2}
\end{figure}

Now, using the $N(1520)$, $N(1700)$, $N(1875)$ and $N(2120)$ nucleon family, the CE of each one of these nucleons with $J^P=\frac32^-$ is computed. Therefore, when  the CE is  interpolated as a function of the $n$ quantum numbers, a CE Regge trajectory is derived, for the $J^P=\frac32^-$ nucleon family resonances, as a dashed curve in Fig. \ref{kaon2}.
The data in the fourth and fifth columns in Table \ref{CES}, for $n=0,1,2,3$, are better represented by the points in Fig. \ref{kaon2} as a function of $n$. 
\noindent The explicit expression of the CE  Regge trajectory, represented by the interpolation dashed curve in Fig. (\ref{kaon2})  reads
\begin{eqnarray}
 {\rm CE}(n) &=&0.007780 n^3 - 0.058302 n^2 + 0.190589 n \nonumber\\&&-0.003969.
 \label{itp1}
   \end{eqnarray} We opted to use a cubic interpolation, as it is sufficient to delimit within $\sim0.19\%$ the standard deviation. 
    
\begin{figure}[H]
\label{cexm32}
\begin{center}
\includegraphics[scale=0.5]{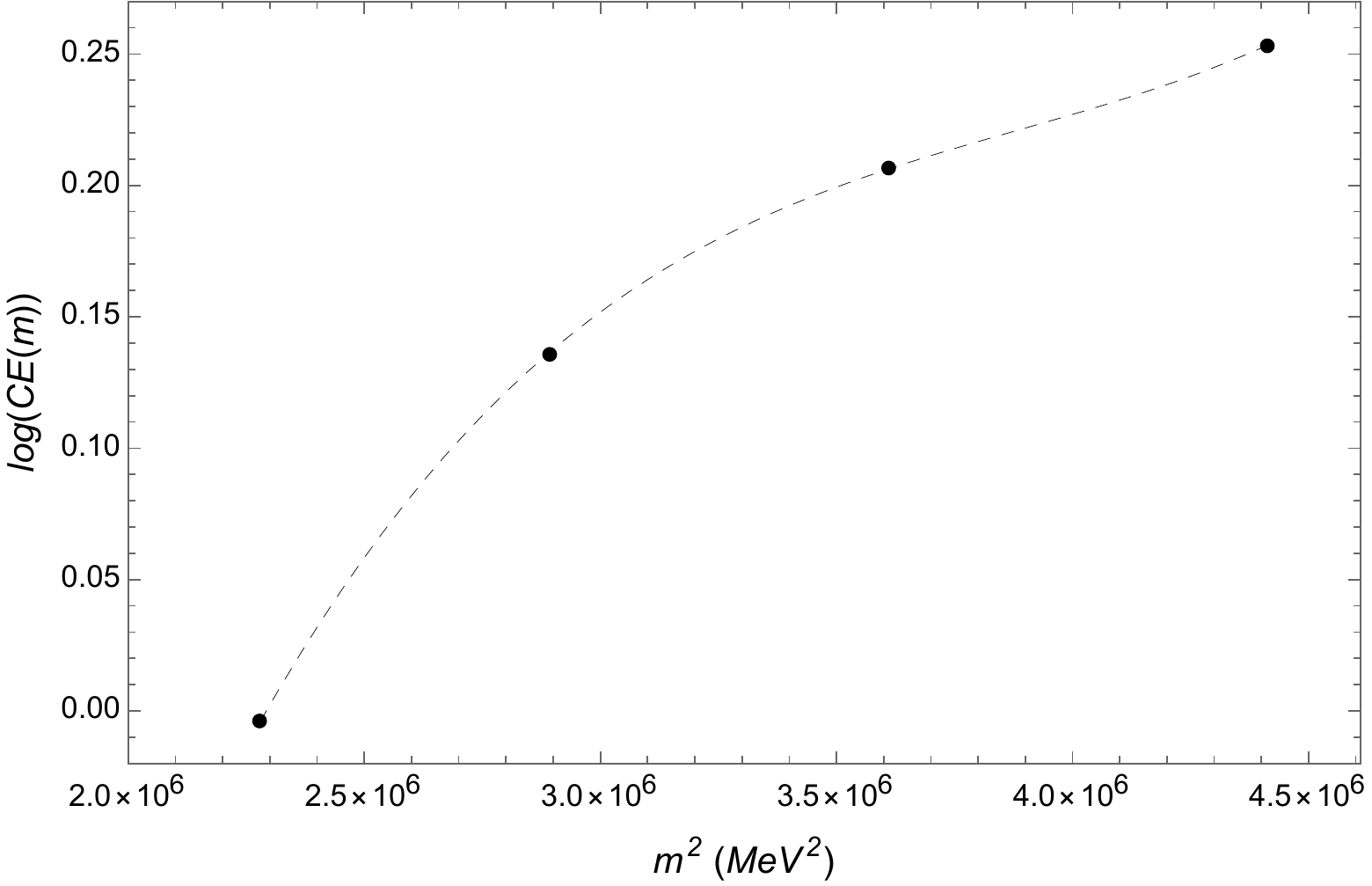}
\end{center}
\caption{ The CE for the nucleon family $J^P=\frac32^-$ information-entropic Regge trajectory, with respect to the squared mass spectra.}\label{f6}
\end{figure}

A second type of configurational-entropic Regge trajectory, relating the CE to (squared) mass spectra of the nucleon family $J^P=\frac32^-$, is shown in Fig. \ref{f6},
 \noindent {\color{black}{The dashed curves in Figs. \ref{kaon2} and \ref{f6}, respectively correspond to 
Eqs. (\ref{itp1}) and (\ref{itq1}).}}
 For the plot in Fig.  \ref{f6}, the configurational-entropic Regge trajectory has the following 
form:
\begin{eqnarray}\label{itq1}
\!\!\!\!\!\!\!\! \log{\rm CE}(m) \!&=\!&\! 3.4700\times 10^{-20} m^6\!\!-4.0410\times 10^{-13} m^4\!\nonumber\\&&1.18206\times 10^{-7}\,m^2+0.84686,    \end{eqnarray} within $0.18\%$ standard deviation. 

 The mass spectra of the $X_{5}, X_{6}$ and $X_{7}$ elements can be easily inferred, by employing Eqs. (\ref{itp1}, \ref{itq1}). In fact, for $n=4$, Eq. (\ref{itp1}) yields 
${\rm CE} = 1.3819$. Then substituting this value in the CE  Regge trajectory (\ref{itq1}), and solving the resulting equation, the solution is the mass 
\beq\label{mx5}
m_{X_5}=2232\pm 11 \;{\rm MeV}.
\eeq  
This nucleon resonance matches the mass and other properties of the $\Xi(2250)$
isospin $\frac12$ xi baryon, whose $J^P$ still needs confirmation
\cite{pdg1}. However, the $\Xi(2250)$ baryon is a potential candidate to represent the $X_5$ nucleon resonance with mass (\ref{mx5}).

Similarly for the next member $X_{6}$ of the spin $\frac12^+$ nucleon family, substituting $n=5$ into Eq. (\ref{itp1}) implies that the corresponding CE has value $1.5903$. Hence, replacing this value into 
the CE Regge trajectory (\ref{itq1}) yields 
\beq\label{mx6}
m_{X_{6}}=2354\pm15\; {\rm MeV}.
\eeq 
Analogous to the previously analyzed case consisting of the $X_5$ nucleon resonance with $J^d=\frac12^+$, also the 
$X_6$ nucleon resonance does match the mass and other properties, like the isospin of the $\Xi(2370)$
 xi baryon \cite{pdg1}. The $\Xi(2370)$ baryon can be a potential candidate to describe the $X_6$ nucleon resonance with mass (\ref{mx6}).

Besides, the same can be accomplished for the $X_{7}$  $J^P=\frac12^+$ nucleon, when Eq. (\ref{itp1}) implies that ${\rm CE}_{X_{7}} = 2.0568$. When replaced into Eq. (\ref{itp1}), it yields the mass 
\beq\label{mx7}
m_{X_{7}}=2473\pm26\; {\rm MeV}. 
\eeq
Again, the derived mass \eqref{mx7} for the $X_7$ nucleon resonance with $J^P=\frac32^-$ matches the mass and other properties, like the isospin of the $\Xi(2500)$
 xi baryon \cite{pdg1}, which may consist of a potential candidate to emulate the $X_7$ nucleon resonance with mass (\ref{mx7}).

\subsection*{Finite Temperature case}

\begin{figure}[h!]
\begin{center}
\includegraphics[scale=0.49]{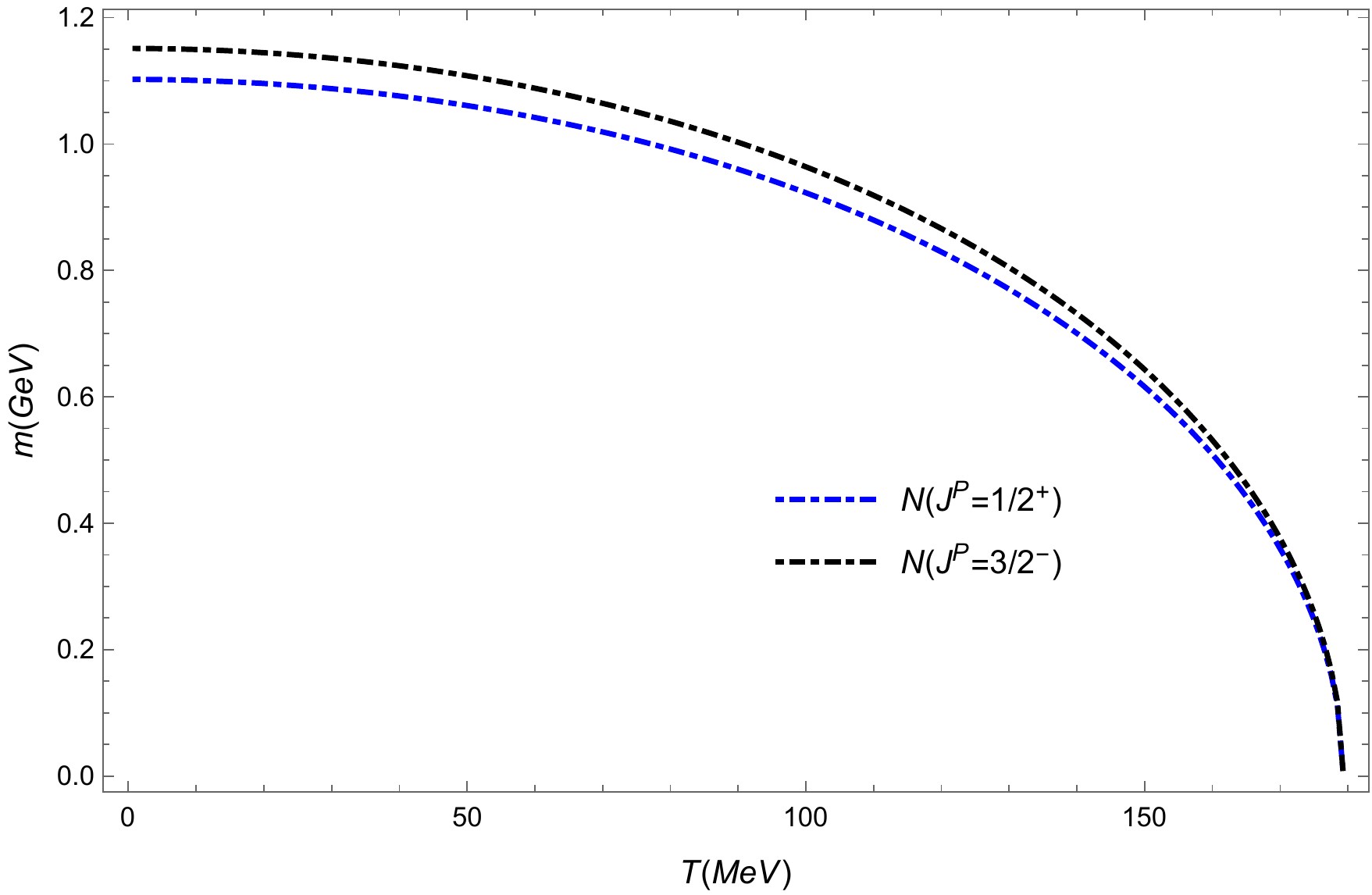}
\end{center}
\caption{ The masses of the ground state $J^P=\frac12^+$ and $J^P=\frac32^-$ as a function of the temperature.}
\label{Mtemp}
\end{figure}

\begin{figure}[h!]
\begin{center}
\includegraphics[scale=0.49]{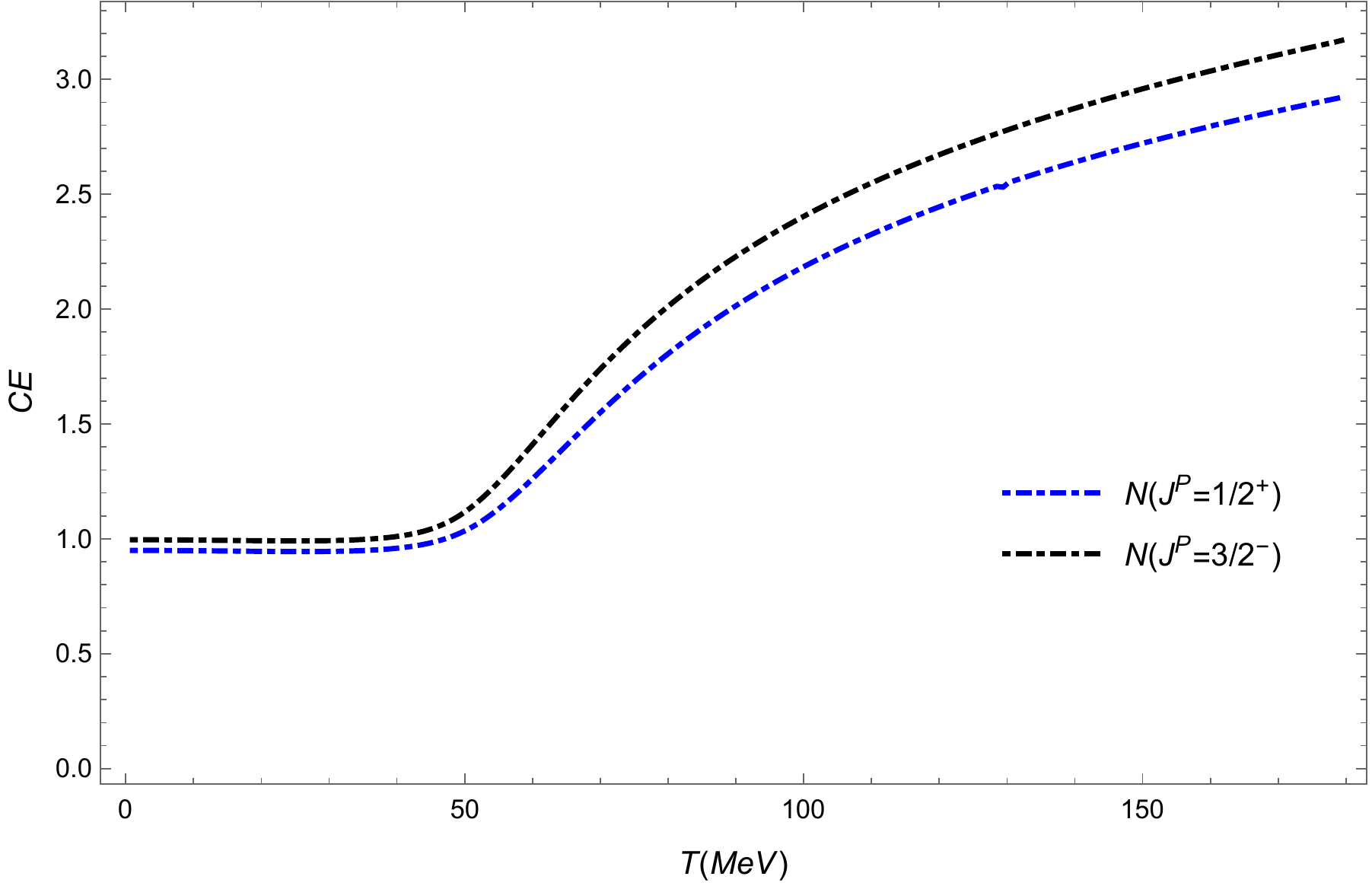}
\end{center}
\caption{ The CE for the ground state  of nucleon resonances  $J^P=\frac12^+$ and $J^P=\frac32^-$   as a function of the temperature.}
\label{CEtemp}
\end{figure}

Applying the same procedure to compute the CE at zero temperature, we evaluate the CE for lightest state of baryons, as a function the temperature.
The solutions of the fields at finite temperature (\ref{solLzero})  and (\ref{solRzero}) are used. The results of CE  for the ground states of $J^P=\frac12^+$ and $J^P=\frac32^-$ nucleon  resonances at finite temperature are presented in Fig. \ref{CEtemp} as a function of the temperature $T$. From Fig. \ref{CEtemp}, one can note that in the range between $0$-$36$ MeV,  the CE is constant. In this range, for $J^P=\frac12^+$, CE $\approx 0.95$, and for $J^P=\frac32^-$, the CE equals 1.0. For higher temperatures,  the CE increases monotonically as a function of the temperature.
  It suggests a decrement in the configurational stability of the nucleon resonances. The baryons are dissociated at  same temperature as can seen in Fig. \ref{Mtemp}  using the relation (\ref{Spectrum}). The same result is evidenced employing  the CE.

\begin{figure}[h!]
\begin{center}
\includegraphics[scale=0.44]{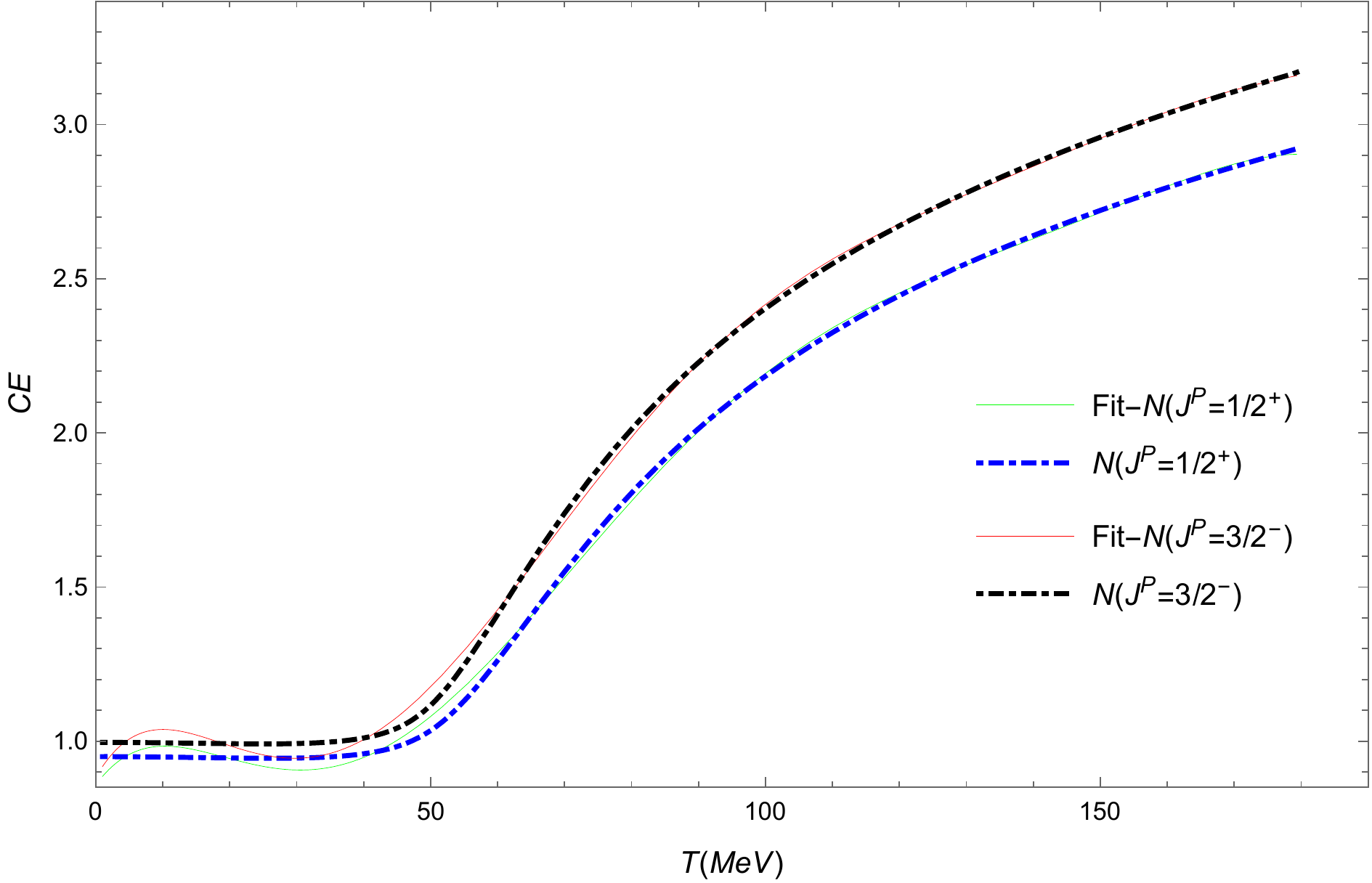}
\end{center}
\caption{ The CE for the nucleon families $J^P=\frac12^+$ and $J^P=\frac32^-$ information-entropic Regge trajectory at finite temperature, with respect to the  mass spectra. }
\label{cefittemp}
\end{figure}

For both the  $J^P=\frac12^+$ and $J^P=\frac32^-$ nucleon families, a scaling law dependence can be implemented. First, Fig. \ref{cefittemp} displays the numerical interpolation for nucleon resonances of $J^P=\frac12^+$, given by
\beq\label{sc1}
{\rm CE}_{\frac12^+}(T) &=&  1.6289\times10^{-14} T^7- 1.4016\times10^{-11} T^6\nonumber\\&&+ 4.7050\times10^{-9} T^5- 
 7.7737\times10^{-7} T^4 \nonumber\\&&+ 6.42109 \times10^{-5}T^3 -2.27622 \times10^{-3}T^2\nonumber\\&&+ 0.0296115 T+
0.858299.
\eeq
 Similarly, numerical interpolation for baryon resonances of $J^P=\frac32^-$ yields 
 \beq\label{sc2}
{\rm CE}_{\frac32^-}(T) &=&  3.21436\times10^{-14} T^7- 2.46643\times10^{-11} T^6\nonumber\\&&+ 7.5303\times10^{-9} T^5- 
 1.1490\times10^{-6} T^4 \nonumber\\&&+ 8.8859 \times10^{-5}T^3 -3.00518 \times10^{-3}T^2\nonumber\\&&+ 0.0377408 T+0.88293,
\eeq
also evincing a scaling law for the CE as a function of the temperature. 
As the CE is constant in the range, $T\lesssim 36$ MeV, the scaling laws  (\ref{sc1}, (\ref{sc2}) hold for $T\gtrsim 60$ MeV within less than 2\% of accuracy.

\section{Concluding remarks}\label{sec4}

In this paper, we computed the CE for the $J^P=1/2^+$ and $J^P=3/2^-$ nucleon families, in the soft wall model for both the zero temperature and finite temperature case. Firstly, we analyzed the zero temperature case using the CE obtained in the holographic model described in Sect.  \ref{sec2}. Two types of configurational-entropic Regge trajectories were derived for each nucleon family, where the CE for nucleon resonances was related to their quantum number $n$ and their mass spectra as well. Through this relation, the next generation of nucleon families was inferred. In the case of the $J^P=\frac12^+$ nucleon family, the derived mass spectra, for the next generation of resonances in this nucleon family corresponding to quantum numbers $n=6,7,8$, were respectively $m_{N_7}=2880\pm 43 \;{\rm MeV}$, $m_{N_{8}}=3062\pm47\; {\rm MeV}.$ and $ m_{N_{9}}=3169\pm51\; {\rm MeV}$. On the other hand, for the $J^P=3/2^-$ nucleon family, the derived mass spectra, for the next generation of resonances, corresponding to quantum numbers $n=4,5,6$,  were  $m_{X_5}=2232\pm 11 \;{\rm MeV}$, $m_{X_{6}}=2354\pm15\; {\rm MeV}$ and $m_{X_{7}}=2473\pm26\; {\rm MeV}$.

    The soft wall AdS/QCD model at small temperatures proposed in Refs. \cite{Gutsche:2019blp,Gutsche:2019pls} for the description of fermionic hadrons was employed to study the dissociation of nucleon families in the context of the CE. The CE  of the ground state for the  $J^P=\frac12^+$ and $J^P=\frac32^-$ nucleon families were analyzed as a function of the temperature, as shown in Fig. (\ref{cefittemp}).  For both cases, the CE monotonically increases with the temperature. It illustrates a decrement in the configurational stability of the system. The result obtained for the CE  is consistent with the result  analyzing the spectrum masses at finite temperature plotted in Fig. (\ref{Mtemp}).  One notices that the mass decreases with increaments in the temperature.  Besides, configurational-entropic Regge trajectories were implemented, relating the CE with the temperature, for  both the  $J^P=\frac12^+$ and $J^P=\frac32^-$ nucleon families.

\paragraph*{Acknowledgments:}  
 LF  is grateful to the National Council for Scientific and Technological Development  -- CNPq (Brazil) under Grant No. 153337/2018-4.
 RdR~is grateful to FAPESP (Grant No.  2017/18897-8), to the National Council for Scientific and Technological Development  -- CNPq (Grants No. 303390/2019-0, No. 406134/2018-9 and No. 303293/2015-2), and to ICTP, for partial financial support. The authors thank to Dr. Allan Gon\c calves for fruitful discussions.

\end{document}